\documentclass[10.5pt,superscriptaddress,longbibliography,twocolumn]{revtex4-2}
\usepackage{amsmath,amssymb,amsfonts,float,graphics,epsfig,epstopdf,color,verbatim,tabularx,bm,multirow,appendix}
\usepackage[utf8]{inputenc}
\usepackage[T1]{fontenc}
\usepackage{xcolor}
\usepackage{lineno}
\usepackage{dsfont}
\usepackage{textcomp}
\usepackage{yfonts}
\usepackage{bm}
\usepackage{subfigure}
\usepackage{mathrsfs}
\usepackage{graphicx}
\usepackage{verbatim}
\usepackage{hyperref}
\usepackage{multirow}
\usepackage{braket}
\usepackage[normalem]{ulem}
\usepackage{tikz}
\usetikzlibrary{calc}
\usetikzlibrary{shapes.multipart}
\usepackage[commandnameprefix=ifneeded,defaultcolor=red]{changes}

\graphicspath{ {images/} }

\DeclareMathOperator{\Tr}{Tr}

\renewcommand{\vec}[1]{{\mathbf #1}}

\newcommand{\comments}[1]{}

\newcommand{\av}[1]{\langle#1\rangle}

\newcommand{\epsb}{\mathcal E^{b}}
\newcommand{\epsf}{\mathcal E^{f}}

\newcommand{\stkout}[1]{\ifmmode\text{\sout{\ensuremath{#1}}}\else\sout{#1}\fi}

\newcommand{\new}[1]{#1}


\begin{document}
 \title{Entanglement Microscopy: \\Tomography and Entanglement in Many-Body Systems}

\author{Ting-Tung Wang}
\altaffiliation{The two authors contributed equally to this work.}
\affiliation{Department of Physics and HK Institute of Quantum Science \& Technology, The University of Hong Kong, Pokfulam Road, Hong Kong SAR}

\author{Menghan Song}
\altaffiliation{The two authors contributed equally to this work.}
\affiliation{Department of Physics and HK Institute of Quantum Science \& Technology, The University of Hong Kong, Pokfulam Road, Hong Kong SAR}

\author{Liuke Lyu}
\affiliation{D\'epartement de Physique, Universit\'e de Montr\'eal, Montr\'eal, QC H3C 3J7, Canada}

\author{William Witczak-Krempa}
\email{w.witczak-krempa@umontreal.ca}
\affiliation{D\'epartement de Physique, Universit\'e de Montr\'eal, Montr\'eal, QC H3C 3J7, Canada}
\affiliation{
 Institut Courtois, Universit\'e de Montr\'eal, Montr\'eal (Qu\'ebec), H2V 0B3, Canada
}
\affiliation{
 Centre de Recherches Math\'ematiques, Universit\'e de Montr\'eal, Montr\'eal, QC, Canada, HC3 3J7
}

\author{Zi Yang Meng}
\email{zymeng@hku.hk}
\affiliation{Department of Physics and HK Institute of Quantum Science \& Technology, The University of Hong Kong, Pokfulam Road, Hong Kong SAR}

\begin{abstract}
\new{We employ a protocol, dubbed entanglement microscopy,
to reveal the multipartite entanglement encoded in the full reduced density matrix of microscopic subregion both in spin and fermionic many-body systems.} 
We exemplify our method by studying the phase diagram near quantum critical points (QCP) in 2 spatial dimensions: the transverse field Ising model and a Gross-Neveu-Yukawa transition of Dirac fermions.
Our main results are: i) the Ising QCP exhibits short-range entanglement with a finite sudden death of the LN 
both in space and temperature; ii) the Gross-Neveu QCP has a power-law decaying fermionic LN consistent with conformal field theory (CFT) exponents; 
iii) going beyond bipartite entanglement, we find no detectable 3-party entanglement with our two witnesses in a large parameter window near the Ising QCP in 2d, in contrast to 1d. \new{We further establish the singular scaling of general multipartite entanglement measures at criticality, and present an explicit analysis in the tripartite case. }
We also analytically obtain the large-temperature power-law scaling of the fermionic LN for general interacting systems. Entanglement microscopy opens a rich window into quantum matter, with countless systems waiting to be explored.
\end{abstract}

\date{\today}
\maketitle

Quantum entanglement can reveal the fundamental organizing principles of quantum matter, ranging from non-Fermi liquids in high-temperature superconductors to black holes~\cite{hartnoll2018holographic,laflorencieQuantum2016}. 
\new{
However, extracting the entanglement structure in realistic many-body systems remains a significant challenge.
A key step toward understanding these systems is reconstructing the quantum state of microscopic subregions, namely those that contain a small number of sites. 
By focusing on smaller subregions, we can fully reconstruct the reduced density matrix (RDM), potentially giving us access to complete entanglement information. 
The entanglement microscopy protocol thus involves first obtaining the RDM numerically, and then calculating multipartite entanglement within these subregions, providing deeper insights into quantum critical systems.
}

\new{This approach complements existing methods that focus on extended ensembles~\cite{Hastings2010measure,humeniukQuantum2012,groverEntanglement2013,albaEntanglement2013,laflorencieQuantum2016,albaOut2017,demidioEntanglement2020,zhaoScaling2022,songExtracting2023,zhangIntegral2023,zhouIncrmental2024}, which handle larger subregions but are often limited in the choice of entanglement measures. 
It also differs from global measures like Quantum Fisher Information (QFI) and Quantum Variance (QV) which capture coarser quantum correlations in the entire system~\cite{Frerot2018, Frerot2019}.} 
In the simplest case of bipartite entanglement, having the full RDM allows us to compute 
the (logarithmic)  negativity~\cite{zyczkowski98,Eisert98,VW02,plenioLogarithmic2005}, an entanglement monotone not polluted by classical correlations. 
However, negativity is challenging to compute in many-body systems for subregions with many degrees of freedom (including quantum field theories) where one can only obtain related quantities via the replica trick~\cite{CCT12,CCT13}. 
These R\'enyi analogs are not entanglement monotones and thus cannot be used to obtain the entanglement structure. This limitation becomes especially clear in boson/spin systems poised near a quantum critical point (QCP), where the logarithmic negativity (LN) decays faster than any power, as shown for CFTs in 1d~\cite{CCT13} and higher $d$~\cite{parez2023fermionic}, while the R\'enyi moments decay algebraically. These are special cases of a general phenomenon regarding the ``fate of entanglement'' \cite{foe} where any multi-partite form of entanglement (including but more general than bipartite) entirely disappears as a system heats up, evolves in time during certain protocols, or as its parts grow separated. 

\new{In the first step of our protocol, we compute the reduced density matrix (RDM) for microscopic or “skeletal” subregions of many-body systems~\cite{Berthiere_2022_Entanglement}. This can be achieved through various computational approaches
such as Exact Diagonalization (ED), Quantum Monte Carlo (QMC), or tensor network approaches such as the Density Matrix Renormalization Group (DMRG). 
In our work, we focus on numerically constructing the RDM using ED and QMC~\cite{mao_sampling_2023}.
 To be more specific, we scrutinize systems near generic 2d quantum phase transitions, including (1) the celebrated QCP in the transverse field Ising model (TFIM) on the square lattice} and (2) a Gross-Neveu-Yukawa \new{(GNY)} QCP of interacting Dirac fermions. Regarding bipartite entanglement, our microscopy yields the LN in the entire phase diagram including at finite temperature with the main findings: i) the Ising model exhibits short-range entanglement with finite sudden death 
as a function of space and temperature; ii) the fermionic variant of the LN, which takes into account the parity superselection rule, shows power-law decay at the \new{GNY} QCP for large separations with exponents consistent with the conformal field theory (CFT) prediction~\cite{parez2023fermionic}.
iii) Going beyond bipartite entanglement, we do not detect any genuine 3-party entanglement in a large window near the 2d Ising QCP, but find its presence in the 1d model both in QMC and with the exact solution via fermionization. Overall, the 2d Ising model shows less bipartite and tripartite entanglement compared to 1d~\cite{osterlohScaling2002,Osborne2002,Giampaolo2013,javanmardSharp2018}.\\

\begin{figure}[htp!]
\includegraphics[width=\columnwidth]{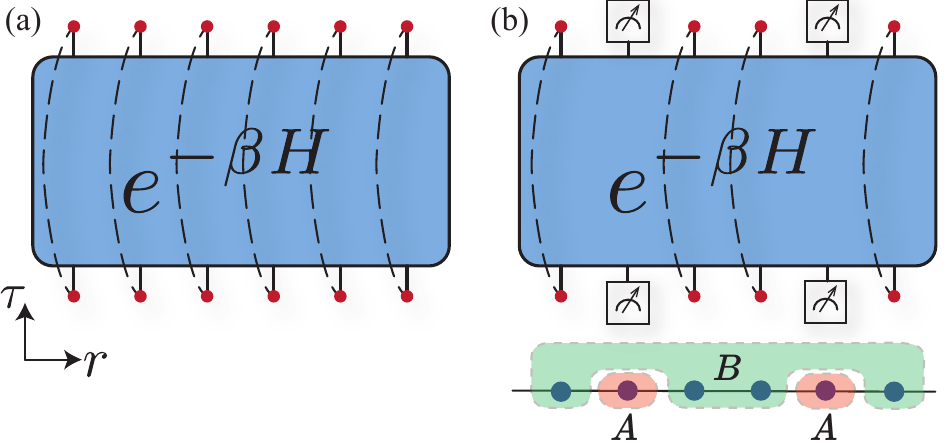}
\caption{\textbf{Sampling the reduced density matrix in path-integral QMC.} 
Panel (a) shows the partition function in the space ($r$) - time ($\tau$) manifold appearing in usual QMC simulations, with the periodic structure of imaginary time. 
Panel (b) shows how the RDM is sampled in such QMC simulations, where one imposes open boundary conditions in time for the skeletal sites in subregion $A$. From the QMC histogram of configurations of these sites, in this case, there will be $4\times 4$ configurations for the 2 spins in $A$, the corresponding matrix elements in the RDM are recorded. The lower panel in (b) demonstrates the spatial partition of $A$ and $B$ in such a setting.}
\label{fig:fig1}
\end{figure}

\noindent{\textbf{\large Results}}\\
\noindent{\textbf{Entanglement microscopy}}

\new{To extract the entanglement information in microscopic subregions for spin models, we use the algorithm introduced in Ref.~\cite{mao_sampling_2023}, schematically shown in Fig.~\ref{fig:fig1}. In general, the matrix elements of the RDM for spins in subregion $A$, $\langle \alpha|\rho_A|\tilde{\alpha}\rangle$ is expressed as}
\begin{equation} \label{eq:rho}
\begin{split}
\langle \alpha|\rho_A|\tilde{\alpha}\rangle&=\new{
\frac{1}{Z}\bra{\alpha}\Tr_B e^{-\beta H}\ket{\tilde{\alpha}}
}\\
&=\new{
\frac{1}{Z}\sum_\xi \bra{\alpha, \xi}e^{-\beta H}\ket{\tilde{\alpha}, \xi},
}
\end{split}
\end{equation}
\new{where $\{|\xi\rangle\}$ forms a complete basis for the environment region $B$ and $\beta$ is the inverse temperature. Eq.~\eqref{eq:rho} can be viewed as a modified partition function with imaginary-time periodicity only in the region $B$~\cite{mao_sampling_2023}, which can be easily simulated in path-integral QMC as shown in Fig.~\ref{fig:fig1}. Please refer to the latter section for a detailed description.}

For interacting fermions, one employs the determinant quantum Monte Carlo (DQMC) scheme~\cite{blankenbeclerMonte1981,scalapinoMonte1981,xuRevealing2019}. In DQMC, the sampling of the RDM is attributed to the fact that fermions are considered free in each auxiliary field configuration, thus the interacting RDM is the weighted average of free RDMs~\cite{Grover_Entanglement_2013}. \new{Recent advances in DQMC have enabled the reconstruction of the RDM for small subregions in many-body quantum systems, such as a four-site plaquette in the two-dimensional fermion Hubbard model~\cite{Humeniuk2019}. 
However, in that work, genuine entanglement between sites within the plaquette has not been studied. Similarly, studies using dynamical mean-field theory (DMFT) on the Kagome Hubbard model have focused on subregions and examined only some diagonal elements of the RDM~\cite{Udagawa2010}, without exploring genuine entanglement.}
For the case that we focus on, where region $A$ contains two sites, the interacting RDM in the occupation-number basis can be constructed with the density correlation function, and fermion Green's function~\cite{Cheong_many_2004}.
We also note that DQMC calculation of R\'enyi negativity with bipartition at finite temperature of interacting fermions, via the replica trick, has recently been worked out~\cite{wangEntanglement2023}. 
However, since the RDM therein is not explicitly computed, only integer moments of LN can be obtained.

\begin{figure*}[htp!]
\centering
\includegraphics[width=\textwidth]{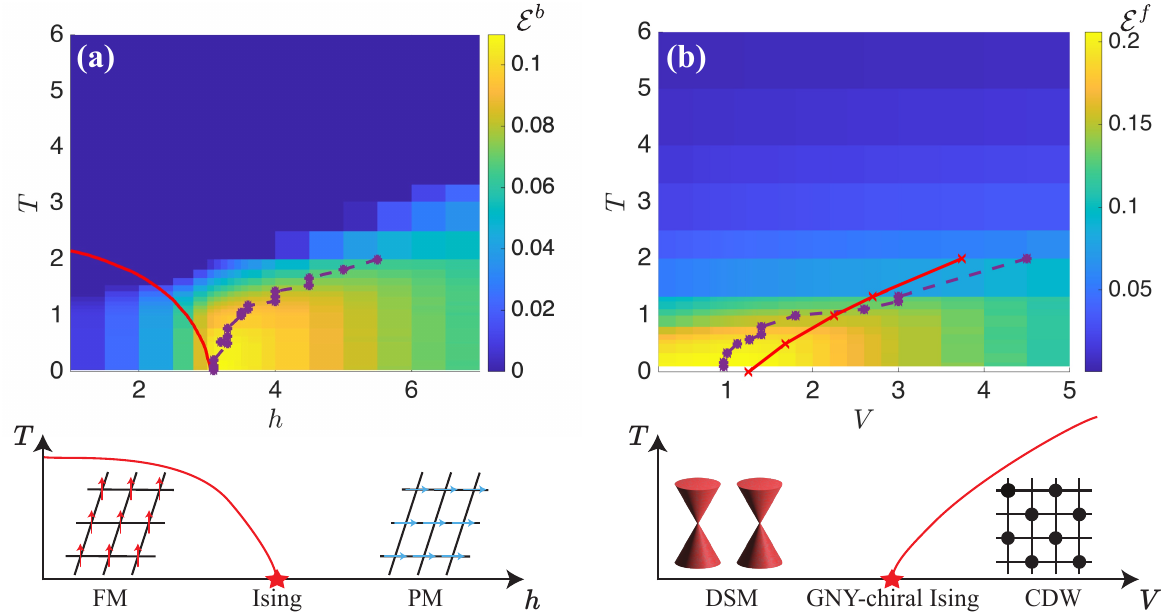}
\caption{\textbf{Logarithmic negativity for adjacent sites.} Color code plots of (a) $\epsb$ in 2d TFIM simulated in a $40\times 40$ square lattice with $\beta$ up to \new{40}, and (b) $\epsf$ in 2d fermion $t-V$ model simulated in a $6\times 6$ square lattice with $\beta$ up to 15.  In both panels, the red curves represent the finite-temperature phase boundaries which end at a zero temperature QCP located at $h_c=3.044$ in 2d TFIM~\cite{hesselmannThermal2016} and $V_c=1.27$ in $t-V$ model. The purple dashed lines show the maxima of $\mathcal E^{b/f}$ for each temperature. In both models, the maxima have converged with system size, as is shown in the SM~\cite{suppl}.  The bottom panel illustrates the schematic phase diagrams.}
\label{fig:fig2}
\end{figure*}

\noindent{\textbf{Bipartite entanglement.}} 
\new{Bipartite entanglement quantifies the quantum correlations between two subsystems, which cannot be reduced to classical correlations. 
Widely used bipartite entanglement measures include entanglement entropy, concurrence, and logarithmic negativity, the latter valued for its simplicity and applicability to both pure and mixed states,
making it a robust tool for studying entanglement in systems near quantum critical points~\cite{Horodecki2009}.
Consider a region $A$, which is divided into two parts, $A_1$ and $A_2$. The logarithmic negativity is defined as a function of the partial transpose on the RDM $\rho_A$ in region $A_1$~\cite{zyczkowski98,Eisert98,VW02}
}
\begin{equation} 
\mathcal E^{b/f}=\ln ||\rho_A^{T^{b/f}_1}||_1
\end{equation} 
where the $|| \cdot ||_1$ is the matrix 1-norm which sums the absolute values of the eigenvalues of the partially transposed matrix $\rho_A^{T^{b/f}_1}$. The bosonic partial transpose of the RDM $\rho_A$ operates as  $\bra{\alpha_{1}\alpha_{2}}\rho_A^{T^b_1}\ket{\bar{\alpha}_{1}\bar{\alpha}_{2}}=\bra{\bar{\alpha}_{1}\alpha_{2}}\rho_A\ket{\alpha_{1}\bar{\alpha}_{2}}$;
the subscript denotes the subregions, $A_1$ or $A_2$,
supporting the states. The superscript $b$ in $\rho_A^{T^b_1}$ refers to the fact that, in doing partial transpose, one simply exchanges two states without any extra factor. For fermionic systems, one can also define the parity-respecting quantity $\epsf$ that takes the phase factor coming from the non-local commutation relation into account, thus recovering the basic properties of the fermionic entanglement measure~\cite{shapourianPartial2017,parez2023fermionic,choi2023finite,Shiozaki_many_2018,Shapourian_Entanglement_2019,wangEntanglement2023}. \\
 
\noindent{\textbf{Tripartite entanglement.}} 
In the above, we have discussed bipartite entanglement between two subregions. Entanglement microscopy can also delve into the much richer realm of multi-party entanglement. We illustrate this by studying tripartite entanglement between 3 sites in the 2d quantum Ising model. In particular, we shall search for genuine tripartite entanglement (GTE), which cannot be described in terms of 2-party entanglement between any two subgroups of spins.  A state with GTE cannot be written as a mixture of bi-separable physical density matrices, $\rho\neq \sum_k p_k \rho_A^k\rho_{BC}^k+ q_k\rho_{AB}^k\rho_C^k+ r_k \rho_{AC}^k\rho_B^k$, where $p_k,q_k,r_k$ are non-negative. A simple example of GTE is given by a pure cat state (also known as GHZ state) of 3 spins, $a|000\rangle+b|111\rangle$, \new{with non-zero $a, b$}. Criteria to detect GTE are more involved than those of the simple bipartite measures such as the LN. 
However, some criteria exist that can detect GTE by establishing that a state is not bi-separable.
Here, we shall focus on two such separability criteria, that we call $W_1$ and $W_2$~\cite{Guhne_2010}. Let us describe $W_1$ ($W_2$ is analogously defined, see Supplemental Material (SM)~\cite{suppl}):
\begin{equation}
    W_1 \!=\! \max_{\,\rm LU} \left(\left|\rho_{1,8}\right|-\sqrt{\rho_{2,2}\rho_{7,7}}-\sqrt{\rho_{3,3}\rho_{6,6}}-\sqrt{\rho_{4,4}\rho_{5,5}} \right)
\end{equation}
where $\rho_{i,j}$ are the matrix element of the 3-site RDM. In order to get a meaningful answer, one has to remove the basis dependence by maximizing among all local unitary (LU) transformations acting on the RDM, \new{$\rho\to U\rho U^\dag$ with $U=U_1\otimes U_2\otimes U_3$}. Being local, such transformations leave the GTE structure invariant. If $W_i >0$, there is GTE, while no conclusion can be reached if $W_i\le 0$. Although not sensitive to all forms of GTE, the $W_i$ criteria do tend to perform well in a variety of states, especially when used together~\cite{Guhne_2010}.\\

\begin{figure*}[htp!]
\centering
\includegraphics[width=\textwidth]{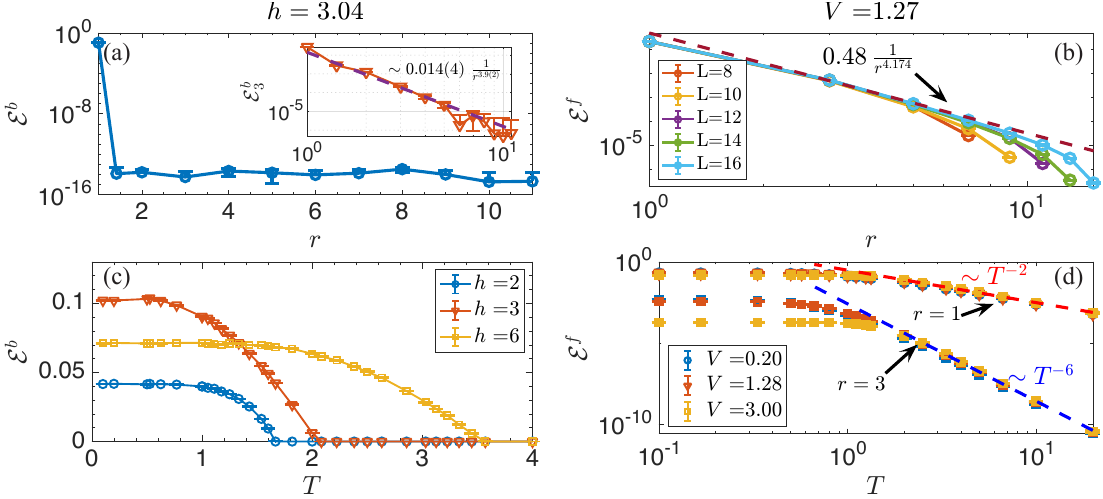}
\caption{\textbf{Spatial and temporal dependence of the logarithmic negativity.} (a) The decay of the two-site $\epsb$ and $\epsb_3$ as a function of separation $r$ at the QCP $h_c=3.04$. The simulation is done with $L=\beta=40$. $\epsb$ exhibits a sudden death beyond one lattice constant, and the $\epsb_3$ exhibits a power-law decay with the fit shown as a purple dashed line; the exponent $3.9(2)$ is consistent with the expected Ising CFT value $4\Delta_\sigma+2\approx4.073$~\cite{parez2023fermionic,Kos_Precision_2016}. (b) The decay of $\epsf$ versus separation $r$ in the $t-V$ model with various system sizes, at the QCP $V_c=1.27$. The simulation is done with $L=\beta$ going from 8 to 16. The blue dashed line shows the algebraic scaling with the power predicted by CFT analysis~\cite{parez2023fermionic,Erramilli-Gross-2023}, with the expected GNY scaling dimension $4\Delta_\psi\approx 4.174$. Panel (c) and (d) demonstrate the temperature dependence of $\epsb$ for the $L=40$ TFIM, and $\epsf$ for the $L=6$ $t-V$ model at various values of $h$ and $V$. In (c) the $\epsb$ always exhibit a finite sudden death temperature and in (d) $\epsf$ decays as $1/T^2$ for adjacent sites $r=1$,
while $1/T^6$ for sites separated by $r=3$ bonds. 
}
\label{fig:fig3}
\end{figure*}

\noindent{\textbf{\large Models}}\\
We perform entanglement microscopy on the 1d and 2d square lattice TFIM, as well as the spinless $\pi$-flux fermion $t-V$ model on a square lattice.
The Hamiltonian of the TFIM reads
\begin{equation}
H=-J\sum_{\langle i,j\rangle}\sigma^z_i \sigma^z_j-h\sum_i \sigma^x_i,
\end{equation}
where we shall set the ferromagnetic exchange coupling to $J=1$; the QCP is at $h=1$ for 1d,  and $h=3.044$ for 2d~\cite{hesselmannThermal2016}. The $T$-$h$ phase diagram is shown in panel (a) of Fig.~\ref{fig:fig2}. As will be shown below, besides the QMC simulation for RDMs, we have also performed exact diagonalization (ED) in 2d for small clusters, to obtain both bipartite and tripartite entanglement measures.

For the spinless fermion $t-V$ model, the Hamiltonian reads,
\begin{equation}
H=-t\sum_{\langle i,j\rangle}(e^{i\theta_{i,j}}c^\dagger_ic_j+ {\rm h.c.})+V\sum_{\langle i,j\rangle}(n_i-\tfrac{1}{2})(n_j-\tfrac{1}{2}),
\end{equation}
it consists of nearest neighbor hopping term with $\pi$-flux and nearest neighbor repulsion. We set $t=1$ as the energy unit. The $\pi$-flux hopping term on the square lattice ($\theta_{i,j}=\frac{\pi}{2}$ on every other column of vertical bonds and 0 otherwise) leads to a dispersion with two Dirac cones at the $X=(\pi,\pm \frac{\pi}{2})$ points in the Brillouin zone (BZ) taking two sites horizontally as a unit cell. At half filling, a positive $V$ drives the system from a Dirac semi-metal (DSM) to a $\mathbb{Z}_2$-symmetry spontaneously broken charge density wave (CDW) state, where one sublattice acquires a higher density than the other. The QCP is described by the ``chiral Ising'' Gross-Nevey-Yukawa (GNY) theory with global symmetry $O(N_f)^2\rtimes \mathbb{Z}_2$ and in our case $N_f=2$ two-component Dirac fermions. The position of the QCP, the critical exponents of the Gross-Neveu transition, have been systematically worked out using QMC, field-theoretical expansions, and conformal bootstrap, for both $N_f=2$~\cite{Wang_Fermionic_2014,Erramilli-Gross-2023,Iliesiu-Bootstrapping-2018,iliesiuBootstrapping2016,Ihrig-Critical-2018,He-Dynamical-2018}, and other values of $N_f$~\cite{Wang_Quantum_2023,Erramilli-Gross-2023,Iliesiu-Bootstrapping-2018,iliesiuBootstrapping2016,Liu-Designer-2020,Ihrig-Critical-2018,He-Dynamical-2018}. The schematic $T-V$ phase diagram is shown in lower part of the panel (b) in Fig.~\ref{fig:fig2}.

\new{By simulating this model with DQMC, we can probe entanglement near the chiral GNY-Ising QCP at $V_c=1.27$, which we located via finite-size scaling and using the critical exponents from conformal bootstrap~\cite{Erramilli-Gross-2023}. The detailed QMC finite-size data analysis and collapse are shown in the SM~\cite{suppl}.}

\noindent{\textbf{Logarithmic negativity}} 

We present the results for $\epsb$ of the TFIM and $\epsf$ of the $t-V$ model.

\noindent{\emph{Ising model}---} Fig.~\ref{fig:fig2} (a) shows nearest neighbor $\epsb$ at finite temperature for the square lattice TFIM of \new{linear system size $L=40$ with inverse temperature $\beta$ up to 40. 
The $\epsb$ maxima, marked by the purple dashed line, shift from $h>h_c$ at high temperature to $h\approx h_c=3.044$ at $T=1/40$.
These numerical values align with prior studies that compute negativity as a function of $h$ rather than temperature or separation~\cite{Braiorr-Orrs_Weyrauch_Rakov_2019}.
Going beyond the numerical results, we provide an argument for the absence of sudden death at small or large $h$ based on evolution of the RDM in space of states~\cite{foe}. At $h=\infty$, the GS becomes the pure product state where all spins point up. The RDM of two adjacent spins is then also a pure product state, which lies at boundary of all separable (un-entangled) RDMs. The entanglement at finite $h$ is only lost at $h=\infty$. In the opposite limit $h\to 0$, the GS becomes the cat or GHZ state $(|\!\uparrow\cdots\uparrow\rangle+|\!\downarrow\cdots\downarrow\rangle)/\sqrt 2$, leading to a rank-deficient RDM with two eigenvalues being 1/2, and the two others zero. A non-full rank separable state lies at the boundary of all separable states (see \cite{WenKempf} and references therein), so that the RDM becomes separable only when $h$ exactly vanishes. 
} 

At all $h$ values, $\epsb$ has a sudden-death temperature, beyond which $\epsb$ is identically zero corresponding to the dark blue region in the phase diagram; it is also shown in panel (c) of Fig.~\ref{fig:fig3}. The existence of a thermal sudden death follows from the general fact that the infinite temperature state lies well within the continent of separable states, and so entanglement must disappear at a finite temperature corresponding to the point at which the system enters the continent~\cite{foe}.

Let us now examine the fate of entanglement as the 2 sites become progressively more separated.
Fig.~\ref{fig:fig3} (a) shows the $\epsb$ between two sites separated with bond-distance $r=|i-j|$. $\epsb$ drops to 0, within machine precision, even for the next-nearest neighbor separation, $r=\sqrt{2}$. Strikingly, the 2d TFIM possesses the shortest possible range of entanglement for two spins, as it only exists for nearest neighbor, even shorter than the 1d case where the sudden-death distance was found to be $r_{\textrm{sd}}=3$~\cite{osterlohScaling2002,Osborne2002,javanmardSharp2018,suppl}. Such a sudden death separation again follows since the state at large separations lies within the separable continent~\cite{foe}.
\new{
Early 1d TFIM studies~\cite{Osborne2002} similarly show rapid concurrence decay with temperature and distance, consistent with our results in Fig.~\ref{fig:fig3}(a) and (c).
Additionally, Roscilde et al.~\cite{Roscilde2004} used quantum Monte Carlo simulations on another Ising model, confirming that concurrence remains short-ranged at a quantum phase transition.}

We also computed the third order moment $\epsb_3$ at the Ising QCP. The moments of the LN are defined~\cite{CCT12,CCT13} as
$\epsb_n=\ln\frac{\Tr\rho_{A}^n}{\Tr[(\rho_{A}^{T_1})^n]},$
which can be easily computed \new{from the full} RDM, or one can sample $\epsb_n$ by using standard QMC and exploiting the replica trick~\cite{Entanglement_Wu_2020}. The orange line in the inset of Fig.~\ref{fig:fig3} (a) shows the power-law decay of $\epsb_3$ with fitted power 3.9(2), which agrees with the CFT prediction $4\Delta_\sigma+2\approx4.073$~\cite{parez2023fermionic,Kos_Precision_2016}. The power-law decay of $\epsb_3$ is in sharp contrast with the rapid sudden death of $\epsb$, and arises since
the former is not a proper entanglement measure, and is thus polluted by non-entangling correlations.\\

\noindent{\emph{Fermion $t-V$ model.}---}Fig.~\ref{fig:fig2} (b) shows the analogous results for $\epsf$ of two adjacent sites for the square lattice $t-V$ model of linear system size $L=6$ with $\beta$ up to 15. As denoted by the purple dashed line, the maxima starts in the ordered CDW phase at high temperature and converges to $V\approx 1$ at low temperature, which is \new{near the QCP but} in the DSM phase. It roughly follows the finite-temperature phase boundary, obtained from finite-size scaling~\cite{suppl} and indicated by the red line, but clearly differs from it at most temperatures. \new{We observe that the nearest-neighbor fermionic LN is less sensitive to the QCP compared to the bosonic LN in the square lattice Ising model. A key difference is that the DSM phase has gapless Dirac fermions, and these lead to a large $\epsf$. The QCP, being driven by the fluctuations of the bosonic order parameter, thus leaves a weaker imprint on $\epsf$. }

As temperature increases, the nearest neighbor $\epsf$ does not suffer a sudden death. Instead, it decays algebraically as shown in Fig.~\ref{fig:fig2} (b) 
and Fig.~\ref{fig:fig3} (d). We can actually obtain the scaling analytically in the large $T$ limit for general models and separations.
In order to determine $\epsf$ in a thermal state $\rho=\exp(-\beta H)/\mathcal Z$, we need to find how the fermion Green's function $g_{ij}=\Tr(c^\dag_i c_j\rho)$ decays to zero. We will then have $\epsf=\alpha |g_{ij}|^2+\cdots$, where the coefficient $\alpha$ depends on the filling, and the ellipsis denote subleading terms. By expanding 
$\exp(-\beta H)=\sum_n (\beta H)^n /n!$ we see that the first non-zero contribution to the Green's function will come from $\Tr(H^{n_{r}} c^\dag_i c_j)$ such that the hopping term of $H^{n_{r}}$ connects sites $i$ and $j$. The Green's function thus decays as $1/T^{n_{r}}$, and we have
\begin{align}
    \epsf=A\, T^{-2n_{r}}+\cdots 
\end{align}
To illustrate the result, let us consider a general Hamiltonian that includes nearest neighbor hopping terms, an example being our $t-V$ model.
If the sites are adjacent, $n_{r}=1$ and we get $1/T^2$.
This scaling is in agreement with the result for adjacent regions in general Gaussian (free) fermionic states~\cite{choi2023finite}, but we emphasize that it holds in interacting systems.
If the sites are not adjacent, 
we will get $1/T^{2n_{r}}$ scaling with $n_{r}\geq 1$.
For instance, in the $t-V$ model for
third nearest neighbor sites, $n_{r}=3$, leading to $1/T^6$ scaling, as shown in Fig.~\ref{fig:fig3}(d).
Further information about the fits are given in the SM~\cite{suppl}.
Finally, the analysis above can be generalized to subregions with more sites: by using the locality of the large-temperature state, we infer that the exponent is the minimal one for all combinations of sites $i\in A_1$ and $j\in A_2$. This will yield $1/T^2$ decay for $\epsf$ of adjacent regions in generic interacting models.
Interestingly, a smaller exponent $n=3/2$ was observed in topological integer quantum Hall states for adjacent regions~\cite{Hall}, which does not contradict our general analysis since quantum Hall states live in the continuum.


Finally, Fig.~\ref{fig:fig3} (b) shows the scaling of $\epsf$ as we increase the separation between the two sites at the GNY QCP. In contrast to $\epsb$ for the Ising QCP, the fermionic LN shows long-range entanglement, manifested by its power-law decay with $r$. The dashed line shows the CFT prediction~\cite{parez2023fermionic} with power $4\Delta_\psi\approx 4.174$, where the fermion scaling dimension was obtained via conformal bootstrap~\cite{Erramilli-Gross-2023}. The algebraic tail of the fermionic LN follows from the general structure of the space of separable states in fermion systems owing to the \new{fermion} parity superselection rule~\cite{foe}.\\

\noindent{\textbf{Genuine Tripartite Entanglement}} 

\begin{figure}[htp!]
\centering
\includegraphics[width=\columnwidth]{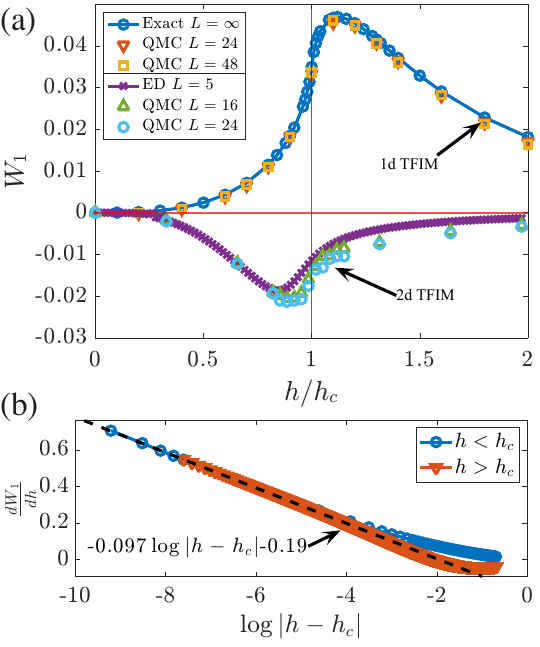}
\caption{\textbf{Searching for genuine tripartite entanglement in the 1d/2d quantum Ising models.} (a) In 1d, the criterion $W_1 \ge0$ detects GTE for 3 consecutive spins at all the values of $h$. In 2d for 3 spins in a square plaquette, $W_1 \le 0$, both in the small size exact diagonalization (ED) and large-scale QMC results, signifying no detection of GTE in a large region near the QCP. Similar behavior for $W_2$ is shown in the SM~\cite{suppl}. All QMC calculations are done with $\beta=L$. \new{Panel (b) shows the logarithmic divergence of the derivative $\frac{dW_1}{dh}$ near QCP for the 1d $L\!=\!\infty$ data: $\frac{dW_1}{dh}|_h= \frac{W_1(h+\Delta h)-W_1(h-\Delta h)}{2\Delta h}$, with step size $\Delta h=10^{-4}$.}}
\label{fig:fig4}
\end{figure}

In 1d, GTE of three consecutive spins was studied in the 1d anisotropic XY model, which includes the Ising case as a special limit~\cite{Giampaolo2013}, using a quantum $k$-separability criterion~\cite{Gabriel2010}. GTE has also been studied
in a cluster-Ising model that gives rise to RDMs with special fine-tuned properties~\cite{giampaoloGenuine2014},
while multi-partite entanglement of the entire system, not subregions as considered here, was studied for small system sizes in the TFIM in dimensions 1 to 3~\cite{dengDetect2010}. 
GTE beyond 1d has not been previously computed.
Fig.~\ref{fig:fig4} shows the GTE criterion $W_1$ for the three closest sites as a function of the transverse field $h$ for both the 1d and 2d TFIMs. 
The 1d data, obtained from the exact RDM via a Jordan-Wigner transformation~\cite{Fagotti2013}, shows that GTE exists for all finite values $h$, and becomes maximal in the paramagnetic phase near the QCP. \new{The absence of sudden death with $h$, can be explained similarly to the LN case above. The asymptotic state at  $h=\infty$ is a pure product state, and so is the RDM of 3 spins. This state actually lies at the boundary of the all biseparable states since we can perturb with an arbitrarily small $\epsilon$: $\sqrt{1-\epsilon}|\uparrow\uparrow\uparrow\rangle+\sqrt{\epsilon}|\downarrow\downarrow\downarrow\rangle$ to generate a cat / GHZ state (with unequal amplitudes), which possesses GTE as detected by $W_1$. At small $h$, we have a rank 2 RDM (since the full GS is the ferromagnetic cat state), which is rank deficient. In the zero-rank subspace, we can perturb with a $W$-state: $(1-\epsilon)\rho+ \epsilon |W\rangle\langle W|$ where $|W\rangle=(|\downarrow\uparrow\uparrow\rangle+ |\uparrow\downarrow\uparrow\rangle+|\uparrow\uparrow\downarrow\rangle)/\sqrt 3$. The $W_2$ criterion detects GTE when $\epsilon>0$. This shows that the asymptotic state is on the boundary of biseparable states, and so the entanglement is only lost at $h=\infty$.   }

In sharp contrast, in the 2d case, the large-scale QMC data has $W_1\leq 0$ for all values $h$, indicating that no GTE is detected. We have also evaluated another complementary criterion $W_2$, and reached the same conclusion in the range $0<h/h_c<2$ (see SM~\cite{suppl}), although more powerful criteria are needed to make a definitive conclusion. Furthermore, we have checked that in both 1d and 2d no GTE is detected if the sites are non-adjacent. Extrapolating to higher dimensions, we expect that GTE decreases with increasing $d$ partly due to the monogamy of entanglement. One could test this in the 3d TFIM on the cubic lattice.

Interestingly, in 2d, using ED on clusters up to 25 sites, we found that a small $W_2\sim 2\times 10^{-4}$ appears at large values of the transverse field $h\gtrsim 8$, and our finite-size scaling analysis suggest that it will survive in the thermodynamic limit (see SM~\cite{suppl}).

\noindent\new{{\bf Scaling of multipartite entanglement near criticality}

Let us now look more closely at the scaling of entanglement measures near the transition. We shall first present an analysis that establishes the singular scaling of $W_1$ in a large class of quantum critical points that includes the Ising CFT in $d\geq 1$, and then give a general argument for other multipartite  entanglement measures (which includes bipartite as a special case). 
Let $h$ be the non-thermal parameter that tunes the system to a QCP.
Let $U_c$ be the optimal LU transformation that yields $W_1$ at $h_c$, and $\check\rho=U_c \rho U_c^\dag$ the transformed RDM. 
As $h$ is varied away from $h_c$, by the Envelope Theorem~\cite{Milgrom2002}, the only contribution comes from the direct variation of the RDM, with $U_c$ unchanged, see Supplemental Material~\cite{suppl} for a derivation. 
The density matrix elements are linear combinations of $n$-point functions of Pauli operators with $n\leq 3$. The Pauli strings with non-zero expectation value need to preserve the symmetries of the finite-size Hamiltonian, which we take to have a unique groundstate. These operators will generically have an overlap with the relevant low energy scalar of the quantum critical theory, which we call $\varepsilon$, with (spatial) scaling dimension $\Delta_{\varepsilon}=d+z-1/\nu$, where $z$ is the dynamical critical exponent, and $\nu$ is the correlation length exponent.
As such, the dominant variation of the expectation value at small $|h-h_c|$ will be proportional to $|h-h_c|^{\Delta_\varepsilon\nu}$.
In turn, this will lead to a singular variation  $W_1= W_1(h_c)+a|h-h_c|^{\Delta_\varepsilon\nu}+\cdots$. 

A more basic argument can be made to explain the singularity 
of a generic entanglement measure or criterion, which we call $\mathcal M$, irrespective of the number of spins / parties considered.
$\mathcal M$ depends on the expectation values within the subregion, and in a symmetric groundstate the relevant scalar will typically dominate those, yielding the singular derivative:
\begin{equation} \label{eq:general-div}
    \frac{d\mathcal M}{dh}\Big|_{h\to h_c} = \alpha\, |h-h_c|^{\Delta_\varepsilon\nu-1}
\end{equation}
It may happen that $\alpha$ vanishes so that the singular behavior appears in higher-order derivatives, as is the case for the next-nearest neighbor concurrence in the 1d TFIM~\cite{osterlohScaling2002}, but that situation is not generic.

In the case of the 1d TFIM, described by the 1d Ising CFT, we have $\Delta_\varepsilon=\nu=z=1$, and a logarithmic divergence appears instead:
\begin{equation} \label{eq:1d-log}
    \frac{dW_1}{dh}\Big|_{h\to h_c} = -\alpha \ln|h-h_c|
\end{equation}
where $\alpha \approx 0.097$, as can be seen in Fig.4b. The GS correlation functions have indeed a logarithmic $\ln|h-h_c|$, as can be verified from the exact solution~\cite{pfeuty}.
The general result \eqref{eq:general-div} also explains the logarithmic divergence numerically seen in various genuine 3-spin and 4-spin entanglement metrics in the 1d TFIM~\cite{Giampaolo2013,Hofmann2014}.
In the 2d Ising CFT, $\Delta_\varepsilon \nu=0.890$, such that the exponent in \eqref{eq:general-div} is $-0.11$.
Such a divergence is difficult to reveal using our finite-size data, but the derivative becomes maximal in close proximity to the QCP, see Fig.~S5 in the Supplemental Materials~\cite{suppl}. 

It is worth noting that in cases of symmetry breaking, which can occur for instance in the magnetic phase of the quantum Ising model in the thermodynamic limit, the critical exponent in the ordered phase will generically differ from that in the symmetric phase.
In the symmetry broken phase,  $\varepsilon$ should then be replaced by the lowest operator that acquires an expectation value. In the Ising case, this will be the order parameter $\phi$, so \eqref{eq:general-div} will depend on $\Delta_\phi$ instead of $\Delta_\varepsilon$ when $h<h_c$. }
\\

\noindent{\textbf{\large Discussion}}\\
\new{In this work, we performed entanglement microscopy to access the true entanglement hidden in quantum many-body systems, by scrutinizing the full RDMs of microscopic subregions obtained by ED and QMC~\cite{mao_sampling_2023}.} We have studied the phase diagram of two representative QCPs in 1 and 2 spatial dimensions: the transverse field Ising model, both in 1d and 2d, and the 2d GNY chiral-Ising transition of Dirac fermions. We found i) the Ising QCP exhibits short-range entanglement with a finite sudden death as a function of separation and temperature; ii) the fermionic QCP exhibits power-law decay of the fermionic LN at large separations between the subregions; iii) there is no detectable 3-party entanglement with our witnesses in a large window near the 2d Ising QCP, while it is present in the 1d counterpart. \new{Finally, we present general non-pertubative results regarding the scaling of multipartite entanglement measures near quantum critical points. Our results demonstrate that precise entanglement microscopy is possible on large-scale quantum many-body systems beyond 1d.} A new window into quantum many-body entanglement is opening, with countless systems waiting to be explored. 

In the future, it will be interesting to explore the multipartite entanglement at various exotic QCPs, including transitions into a non-Fermi-liquid~\cite{jiangMany2023,liuItinerant2019}, deconfined QCPs for both fermion and spin models~\cite{zhaoScaling2022,liaoTeaching2023,liuFermion2023,songDeconfined2023,songExtracting2023,dengDiagnosing2024,demidioEntanglement2024}, and symmetric mass generation transitions~\cite{liuDisorder2024}. \\

\noindent{\textbf{\large Methods}}\\
\noindent{\textbf{Stochastic series expansion.}} 
\new{In the SSE simulation~\cite{sandvik2019stochastic}, the partition function $Z$ is Taylor-expanded as a power series of Hamiltonian operators, and each term is sampled stochastically. As mentioned in the previous section, the RDM matrix element resembles a partition function but with imaginary time boundary conditions open in the region $A$, which is expressed as
\begin{equation}
\begin{split}
\langle \alpha|\rho_A|\tilde{\alpha}\rangle&=
\new{
\frac{1}{Z}\sum_\xi \bra{\alpha, \xi}e^{-\beta H}\ket{\tilde{\alpha}, \xi}
}\\
&=\new{
\frac{1}{Z}\sum_\xi\sum_n\frac{\beta^n}{n!}\bra{\alpha, \xi}H^n\ket{\tilde{\alpha}, \xi}
}
\end{split}
\end{equation}
after series expansion of $e^{-\beta H}$. Therefore, the sampling is almost identical to the standard SSE, but now we simply break the imaginary-time periodicity for loops that pass through sites in the region $A$ when doing loop updates. The frequency of any configuration pair $(\alpha,\tilde{\alpha})$ is proportional to the magnitude of the corresponding element in the RDM, i.e., $\bra{\alpha}\rho_A \ket{\tilde{\alpha}}$, according to the detailed balance condition. That is,

\begin{equation} \label{eq:SSE_RDM}
\frac{\bra{\alpha}\rho_A\ket{\tilde{\alpha}}}{\bra{\alpha'}\rho_A\ket{\tilde{\alpha}'}}=\frac{P(\alpha,\tilde{\alpha})}{P(\alpha',\tilde{\alpha}')},
\end{equation} 
with $P$ denotes the probability of pair of states $(\alpha,\tilde{\alpha})$ to appear. Interested readers are welcome to read Ref.~\cite{mao_sampling_2023} for more details.}

\noindent{\textbf{Determinant quantum Monte Carlo.}} 
For the $t-V$ model, we employ the determinant quantum Monte Carlo (DQMC) scheme~\cite{blankenbeclerMonte1981,scalapinoMonte1981,xuRevealing2019}, where the fermion interaction is decoupled into fermion bilinear coupled to the auxiliary field. In each configuration, the free fermionic Green's function can be calculated, and higher-order correlation functions are evaluated using Wick's theorem.\\

\noindent{\textbf{\large Data availability}}\\
The data that support the findings of this study are available from the corresponding author upon reasonable request.\\

\noindent{\textbf{\large Code availability}}\\
All numerical codes in this paper are available upon reasonable request to the authors.\\

\bibliographystyle{longapsrev4-2}
\bibliography{bibtext}

\noindent{\textbf{\large Acknowledgments}}\\
 We acknowledge discussions with Gilles Parez. T.-T.W., M.H.S. and Z.Y.M. acknowledge the support from the
Research Grants Council (RGC) of Hong Kong Special Administrative Region (SAR) of China (Project Nos. 17301721, AoE/P701/20, 17309822, C7037-22GF, 17302223), the ANR/RGC Joint Research Scheme sponsored by RGC of Hong Kong and French
National Research Agency (Project No. A HKU703/22) and the HKU Seed Funding for Strategic Interdisciplinary Research. 
W.W.-K.\/ and L.L.\/ are supported by a grant from the Fondation Courtois, a Chair of the Institut Courtois, a Discovery Grant from NSERC, and a Canada Research Chair.
We thank HPC2021 system under the Information Technology Services and the Blackbody HPC system at the Department of Physics, University of Hong Kong, as well as the Beijng PARATERA
Tech CO.,Ltd. (URL: https://cloud.paratera.com) for providing HPC resources that have contributed to the research results reported within this paper.\\

\noindent{\textbf{\large Author contributions}}\\
T.-T.W. and M.H.S. \new{realized} the RDM sampling algorithm. T.-T.W. performed the QMC simulation for fermion $t-V$ model. M.H.S. performed the QMC simulation for TFIM. L.L. calculated the exact result in 1d TFIM and performed the ED calculation in 2d TFIM. All the authors contributed to the analysis of the results.
W.W.-K. and Z.Y.M. supervised the project.\\

\noindent{\textbf{\large Additional information}}\\
\noindent
\textbf{Supplementary Information} is available in the online version of the paper.\\

\noindent
\textbf{Competing interests:} The authors declare no competing interests.\\

\clearpage
\onecolumngrid

\begin{center}
	\textbf{\large Supplemental Material for \\"Entanglement Microscopy: Tomography and Entanglement in Many-Body Systems"}
\end{center}
\setcounter{equation}{0}
\setcounter{figure}{0}
\setcounter{table}{0}
\setcounter{page}{1}
\setcounter{section}{0}

\makeatletter
\renewcommand{\theequation}{S\arabic{equation}}
\renewcommand{\thefigure}{S\arabic{figure}}
\setcounter{secnumdepth}{3}

In this supplementary material, we will provide some additional data on 1d and 2d TFIM in Sec.~\ref{sec:SM1}, and 2d fermion $t-V$ model in Sec.~\ref{sec:SM2}.

In Sec.~\ref{sec:SM1}, we will show the 1d result on the LN $\epsb$ and systematic comparison with ED in addition to the 2d result, which has been covered in the main text. We have also done a finite size convergence test in several parameter settings.

In Sec.~\ref{sec:SM2}, we will show some observables we collected in the $t-V$ model with $L=\beta$. Finite-size scaling, single-particle gap scaling and Green's function decay are done to justify the location of QCP. LN in this setting is also computed and shown. Then, we will show the finite-temperature data including a similar convergence test and finite size scaling to locate the finite-temperature CPs. Lastly, a color code plot of $\epsb$ is presented.

\section*{1d \& 2d Transverse Field Ising Model}  
\label{sec:SM1}
\begin{figure}[htp!]
\includegraphics[width=\columnwidth]{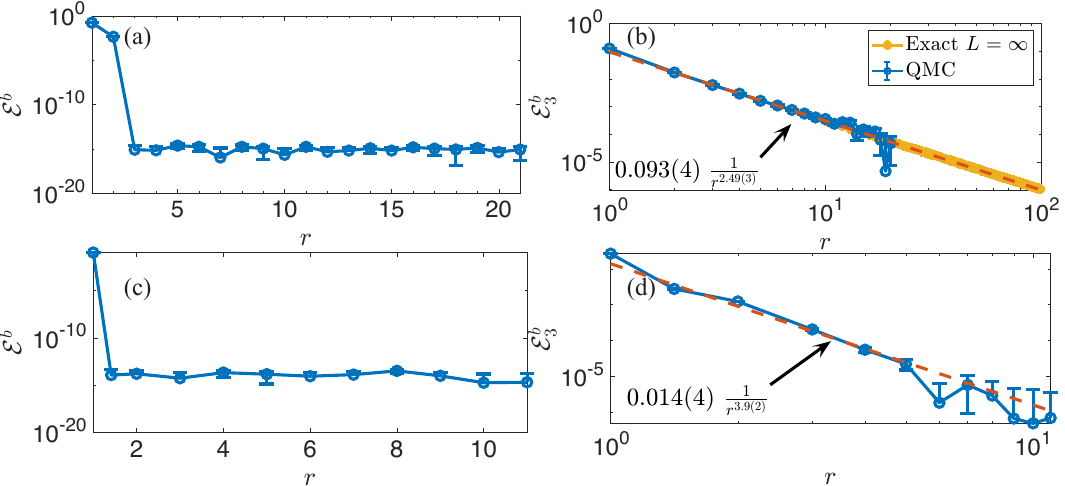}
\caption{\textbf{2-site entanglement negativity with separation $r$ of 1d and 2d transverse field Ising models.} Log negativity $\epsb$ (a), and its third order moment $\epsb_3$ (b) in 1d Ising model with system size and inverse temperature $L=\beta=128$, and $h=1$. Log negativity $\epsb$ (c), and its third order moment $\epsb_3$ (d) in 2d Ising model with system size and inverse temperature $L=\beta=40$, and $h=3.044$.}
\label{fig:fig_1sing}
\end{figure}

Fig.~\ref{fig:fig_1sing} (a) and (b) show the decay of $\epsb$ and $\epsb_3$ against the separation distance between two sites for the (1+1)d Ising model on the QCP. The logarithmic negativity (LN) $\epsb$ drops abruptly to $10^{-16}$ (machine precision) with sudden-death distance $r_{\textrm{sd}}=3$ \cite{javanmardSharp2018}, while its third moment decay algebraically with fitted power $2.49(3)$, which agrees with both exact calculation using Jordan Wigner Transformation and CFT prediction $2+4\Delta_\sigma=2.5$~\cite{parez2023fermionic}.

Similar to the case in 1d TFIM, as shown in Fig.~\ref{fig:fig_1sing} (c), the LN $\epsb$ drop to the magnitude of machine error with sudden-death distance $r_{\textrm{sd}}=\sqrt{2}$, i.e., only nearest neighbor entangle exists. Its third order moment $\epsb_3$ decay algebraically with fitted power $3.9(2)$, where CFT prediction is $4\Delta_\sigma+2\approx4.073$~\cite{parez2023fermionic,Kos_Precision_2016}.

\begin{figure}[htp!]
\includegraphics[width=\columnwidth]{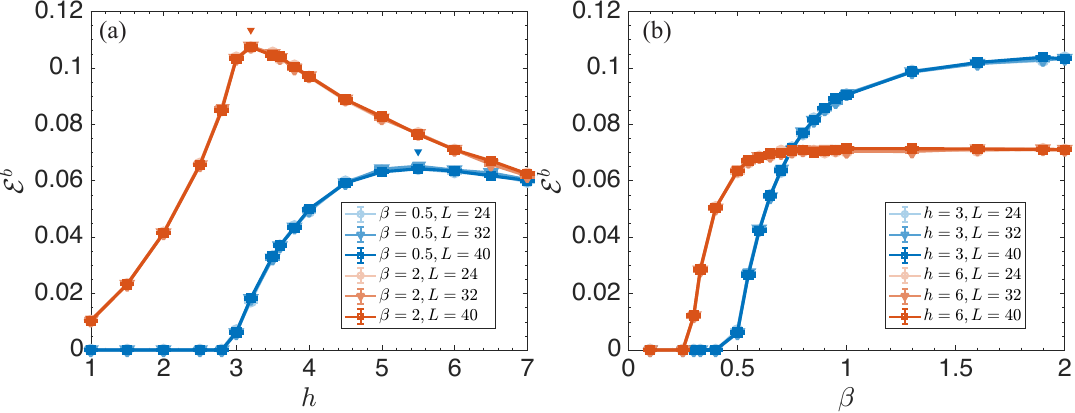}
\caption{\textbf{Nearest neighbor LN $\epsb$ at various system sizes.} (a) $\epsb$ changes against the transverse field $h$ at different temperature and sizes. For each temperature, $\epsb$ is converged against system size. The triangles indicate the maxima of $\epsb$ along the $h$ axis (the purple dots in Fig.~\ref{fig:fig2} in the main text). (b) $\epsb$ changes against the inverse temperature $\beta$ at different transverse field $h$ and sizes. For each $h$, $\epsb$ is also converged against system size.}
\label{fig:fig_Ising_size}
\end{figure}

Fig.~\ref{fig:fig_Ising_size} shows the convergence test of the nearest neighbor LN $\epsb$ against system size $L$. In all cases, the results from these three sizes show good convergence in terms of size. This strongly suggests that the $L=40$ data presented in main text has negligible finite-size effect.

\subsection*{Comparison with Exact Diagonalisation}

\begin{figure}[htp!]
\includegraphics[width=\columnwidth]{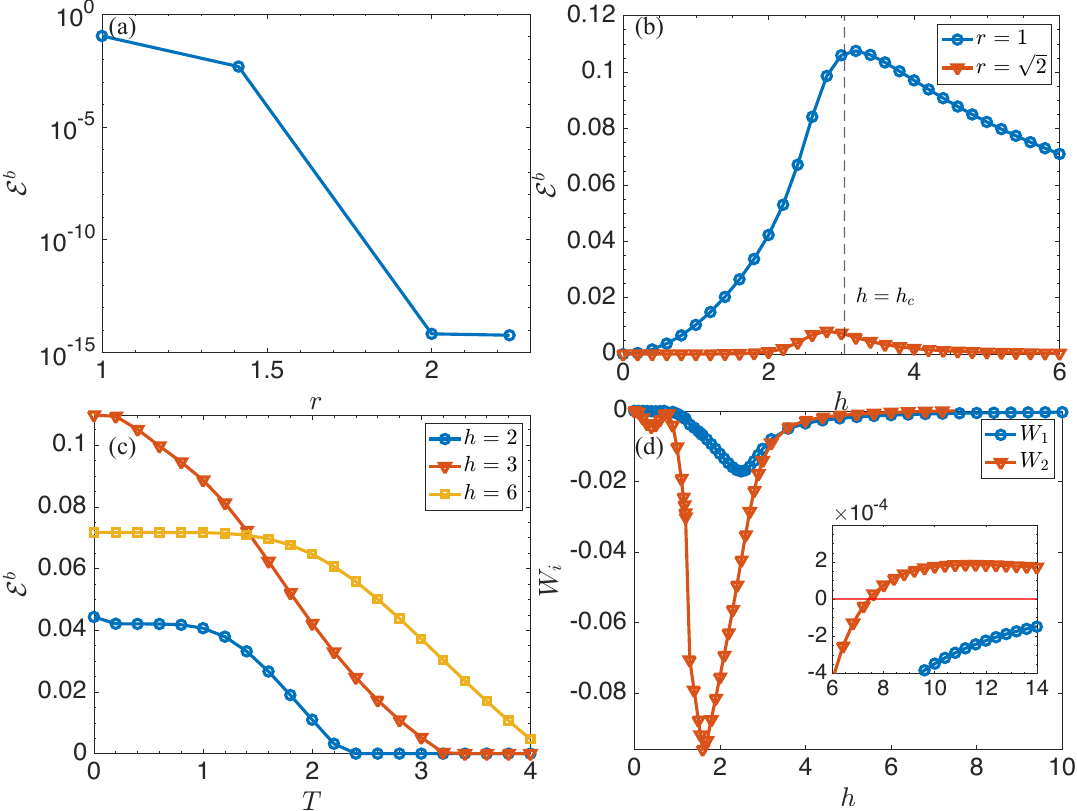}
\caption{\textbf{Entanglement measures from ED.} (a) The decay of the two-site $\mathcal{E}^{b}$ against the separation distance $r$, with $L_{x},L_{y}=5,5$, $T=0$ and $h=3.04$ (b) Two-site $\mathcal{E}^{b}$ against transverse field h for two sites at distance $r=1$ and $\sqrt{2}$, with $L_{x},L_{y}=4,5, T=0$. (c) The decay of two-site $\mathcal{E}^{b}$ with temperature at distance $r=1$ for $h=2,3,6$, with $L_{x},L_{y}=4,4$, $T=0$. (d) Two GTE criteria against transverse field $h$, with $L_{x},L_{y}=4,5$, $T=0$. The three sites form the smallest right-triangle. The inset show $W_2$ going negative for large $h$. }
\label{fig:ed}
\end{figure}

\begin{figure}[htp!]
\includegraphics[width=0.5\columnwidth]{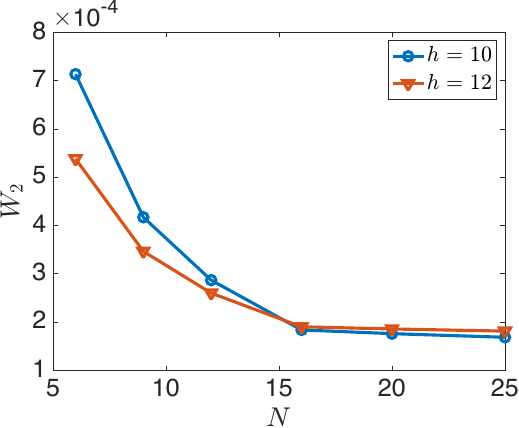}
\caption{\textbf{GTE at large $h$ scaling with size.} GTE criterion $W_2$ has a small positive value of the order $10^{-4}$ at large $h$. We show how this value scales with the size of the system, where $N=L_x L_y$ is the number of spins. For $N=5\times 5$, $W_2$ appears to have converged.} 
\label{fig:ed-size-dependence}
\end{figure}

For small system sizes, we can obtain RDM through exact diagonalisation (ED) and compare with the QMC results. In this section, we compute different entanglement measures using ED. Apart from some finite size effect, these results agree with that of QMC. 
\subsubsection*{$T=0$}
At $T=0$, RDM is constructed from the ground state $\psi_0$ only: $\rho_A = \Tr_B\{\vert \psi_0 \rangle \langle \psi_0 \vert\}$. In Fig.~\ref{fig:ed}~(a),(b),(d), we show how the two-site $\mathcal{E}^{b}$ change with distance $r$ and field $h$, and how the GTE critera $W_1$, $W_2$ change with field $h$.

Fig.~\ref{fig:ed}~(a) shows the decay of the two-site $\mathcal{E}^{b}$ against the separation distance $r$, which can be compared with the $L=40$ QMC result in Fig.~\ref{fig:ed}~(a). We see that $\ensuremath{\mathcal{E}^{b}(1)}\approx0.11$ in both cases, but the ED $\mathcal{E}^{b}(r)$ drop to zero (within machine precision) at distance $r=2$, while the QMC result drops to zero already at distance $r=\sqrt{2}$. This is likely a finite size effect, as with ED we are limited to $L=5$. 

Fig.~\ref{fig:ed}~(b) shows the two-site $\mathcal{E}^{b}$ as a function of the transverse field $h$. At both distance $r=1$ and $r=\sqrt{2}$, $\mathcal{E}^{b}$ peak at around $h=3$, close to the QCP. At large $h$, $\mathcal{E}^{b}(r)$ decays algebraically for $r=1$ and $\sqrt 2$. 

Fig.~\ref{fig:ed}~(d) shows two GTE criteria $W_{1},W_{2}$ against $h$. The two criteria are defined in Ref.~\cite{Guhne_2010} as 
$$\begin{aligned}
W_{1}[\rho] & = \max_{\,\rm LU}\left( \left|\rho_{1,8}\right|-\sqrt{\rho_{2,2}\rho_{7,7}}-\sqrt{\rho_{3,3}\rho_{6,6}}-\sqrt{\rho_{4,4}\rho_{5,5}}
\right)
\\
W_{2}[\rho] & = \max_{\,\rm LU}\left(\left|\rho_{2,3}\right|+\left|\rho_{2,5}\right|+\left|\rho_{3,5}\right|-\frac{1}{2}\left(\rho_{2,2}-\rho_{3,3}+\rho_{5,5}\right)-\sqrt{\rho_{1,1}\rho_{4,4}}-\sqrt{\rho_{1,1}\rho_{6,6}}-\sqrt{\rho_{1,1}\rho_{7,7}}
\right)
\end{aligned}$$
where the two functions are maximized with respect to local unitary transformations $U=U_{1}\otimes U_{2}\otimes U_{3}$ to the reduced density matrix $\rho$. The two separability criteria are defined such that if $W_{i}>0$, there is genuine three-qubit entanglement. If $W_{i}\leq 0$, we cannot conclude anything. $W_{1}$ and $W_{2}$ are strong criteria for GHZ and W states, respectively. In Fig.~\ref{fig:ed}~(d), we see that near $h\sim h_c$ both criteria are negative. For $h>7$, the inset shows that $W_{2}$ go positive for $h>7.6$, indicating a small GTE at large $h$. 
In Fig.~\ref{fig:ed-size-dependence}, we see that $W_2$ seem to converge to a fixed value as we increase the size of the system at large $h$. 
This suggests that the small GTE at large $h$ should survive in the thermodynamic limit. 

\begin{figure}[htp!]
    \centering
    \includegraphics[width=0.6\columnwidth]{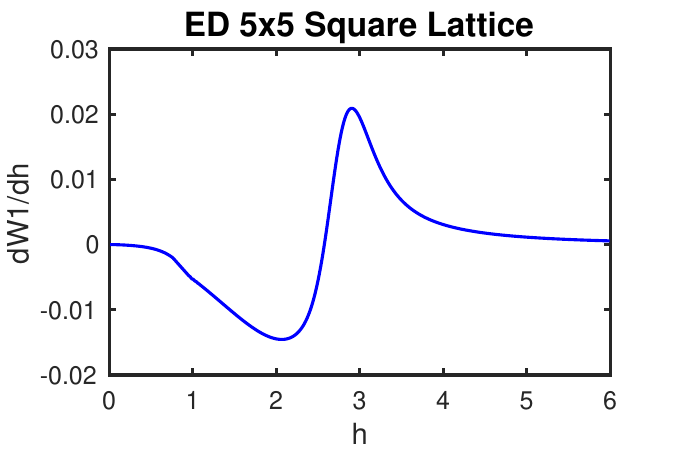}
    \caption{Derivative of $W_1$ with respect to h for the ED 5x5 Square Lattice.}
    \label{fig:dW1_vs_h_derivative_L5_ED}
\end{figure}

Building upon our discussion in the main text, we now present the derivative of $W_1$ with respect to $h$ for the 2D system with $L=5$ using ED, as shown in Fig.~\ref{fig:dW1_vs_h_derivative_L5_ED}. The derivative exhibits a pronounced peak in close proximity to the quantum critical point (QCP), reinforcing the main text's observation that the derivative becomes maximal near the QCP. 

\subsubsection*{Finite T}
At finite temperature $T$, RDM is constructed with the full energy spectrum: $\rho_A = \frac{1}{Z}\Tr_B\{\sum_n e^{-\beta E_n} \vert \psi_n \rangle \langle \psi_n \vert\}$. If $T$ is not high, the Boltzmann factor $e^{-\beta E_n}$ of high energy eigenstates are diminishing, so we can cut the summation to a finite number of lowest energy eigenstates. We found that for $30$ lowest eigenstates, $\mathcal{E}^{b}$ already converges to the values shown in Fig.~\ref{fig:ed}~(c). 

Fig.~\ref{fig:ed}~(c) shows the decay of $\mathcal{E}^{b}(1)$ with temperature, which is very similar to the $L=40$ QMC result in Fig.~\ref{fig:ed}~(c), in terms of the shape of the three curves and the magnitude. The sudden death temperatures are different from that of QMC, likely due to the finite size effect.

\subsection*{Example of RDM}

Here we give an example result of 3-site RDM for the 2d TFIM with $L=\beta=24$ and $h=3$. The three sites are chosen the same way as we did in investigating tripartite entanglement, and form the smallest right triangle with the first site on the right angle. The RDM reads:
$$\begin{pmatrix}
0.22268(5) & 0.13781(3) & 0.13781(3) & 0.09887(3) & 0.12589(4) & 0.11605(3) & 0.11605(3) & 0.12789(5) \\0.13781(3) & 0.09914(2) & 0.09388(4) & 0.08048(2) & 0.08409(2) & 0.09241(2) & 0.08746(3) & 0.11605(3) \\0.13781(3) & 0.09388(4) & 0.09914(2) & 0.08048(2) & 0.08409(2) & 0.08746(3) & 0.09241(2) & 0.11605(3) \\0.09887(3) & 0.08048(2) & 0.08048(2) & 0.07904(3) & 0.06672(3) & 0.08409(2) & 0.08409(2) & 0.12589(4) \\0.12589(4) & 0.08409(2) & 0.08409(2) & 0.06672(3) & 0.07904(3) & 0.08048(2) & 0.08048(2) & 0.09887(3) \\0.11605(3) & 0.09241(2) & 0.08746(3) & 0.08409(2) & 0.08048(2) & 0.09914(2) & 0.09388(4) & 0.13781(3) \\0.11605(3) & 0.08746(3) & 0.09241(2) & 0.08409(2) & 0.08048(2) & 0.09388(4) & 0.09914(2) & 0.13781(3) \\0.12789(5) & 0.11605(3) & 0.11605(3) & 0.12589(4) & 0.09887(3) & 0.13781(3) & 0.13781(3) & 0.22268(5)
\end{pmatrix}$$

\new{
\section*{Derivation of $dW_1/dh$}
In this section, we aim to give a general formula for the derivative of any genuine multipartite entanglement (GME) measure \(W\), which includes both \(W_1\) and \(W_2\), and depends on some optimization procedure. This problem is part of a broader class of problems involving the differentiation of a maximum value function, which is addressed by the well-known Envelop Theorem~\cite{Milgrom2002}. Specifically, we will demonstrate that the derivative \(\frac{d W_1}{d h}\) exhibits a linear dependence on \(\frac{d \rho}{d h}\), achieved by analyzing the behavior of the objective function \(W(h)\) under variations in the parameter \(h\).

Consider \(\rho(h)\) parametrized by a single parameter \(h\). The local unitary transformation \(U = U_1 \otimes U_2 \otimes U_3\) is parametrized by a set of angles \(x = \{x_i\}\). The objective function is defined as \(f(x; h)\). Using this, we define the quantity \(W(h)\) as:

$$
W(h) = \max_x f(x; h)
$$

The optimal set of angles \(x^*\) must satisfy the condition:

$$
\frac{\partial f}{\partial x_i}(x^*) = 0 \quad \forall i
$$

Using these optimal angles \(x^*(h)\), the function \(W(h)\) can be rewritten as:

$$
W(h) = f(x^*(h), h)
$$

Now, taking the derivative of \(W(h)\) with respect to \(h\), we obtain:

$$
\frac{d W}{d h} = \sum_{i} \frac{\partial f(x^*)}{\partial x_i} \frac{\partial x_i^*}{\partial h} + \frac{\partial f}{\partial h}
$$

Since the optimal angles satisfy \(\frac{\partial f}{\partial x_i}(x^*) = 0\), the first term vanishes, leaving us with:

$$
\frac{d W(h)}{d h} = \frac{\partial f(x^*(h), h)}{\partial h}
$$

Next, we apply these results to the specific case of \(W_1\) by defining the function \(g(\rho)\) as:

$$
g(\rho) = \left|\rho_{1,8}\right| - \sqrt{\rho_{2,2} \rho_{7,7}} - \sqrt{\rho_{3,3} \rho_{6,6}} - \sqrt{\rho_{4,4} \rho_{5,5}}
$$

The density matrix \(\tilde{\rho}\) after the unitary transformation \(U(x)\) is given by:

$$
\tilde{\rho} = U(x) \rho(h) U^\dagger(x)
$$

Thus, the objective function for \(W_1\) becomes:

$$
f(x; h) = g(\tilde{\rho})
$$

We now take the derivative of \(W_1(h)\) with respect to \(h\):

$$
\frac{d W_1}{d h} = \frac{\partial g}{\partial h} = \sum_{ij} \frac{\partial g}{\partial \tilde{\rho}_{ij}} \frac{\partial \tilde{\rho}_{ij}}{\partial h}
$$

Since \(\tilde{\rho} = U(x^*) \rho(h) U^\dagger(x^*)\), we can apply the chain rule to write:

$$
\frac{\partial \tilde{\rho}}{\partial h} = U(x^*) \frac{\partial \rho(h)}{\partial h} U^\dagger(x^*)
$$

Substituting this into the expression for \(\frac{d W_1}{d h}\), we get:

$$
\frac{d W_1}{d h} = \sum_{ij} \frac{\partial g}{\partial \tilde{\rho}_{ij}} \left[ U(x^*) \frac{\partial \rho(h)}{\partial h} U^\dagger(x^*) \right]_{ij}
$$

Finally, introducing the optimal unitary transformation \(U_c = U(x^*)\), the result simplifies to:

$$
\frac{d W_1}{d h} = \sum_{ij} \frac{\partial g}{\partial \tilde{\rho}_{ij}} \left[ U_c \frac{\partial \rho(h)}{\partial h} U_c^\dagger \right]_{ij}
$$

At this point, we have arrived at the key result:

$$
\frac{d W_1}{d h} = \sum_{ij} \frac{\partial g}{\partial \rho_{ij}} \left[ U_c \frac{\partial \rho(h)}{\partial h} U_c^\dagger \right]_{ij}
$$

where \(U_c\) is the optimal unitary matrix. We have thus shown that the derivative \(\frac{d W_1}{d h}\) depends linearly on \(\frac{d \rho}{d h}\), as required.
}

\section*{Fermion $t-V$ model}  
\label{sec:SM2}
\subsection*{The $N=4$ GNY quantum critical point}

\begin{figure}[htp!]
\includegraphics[width=\columnwidth]{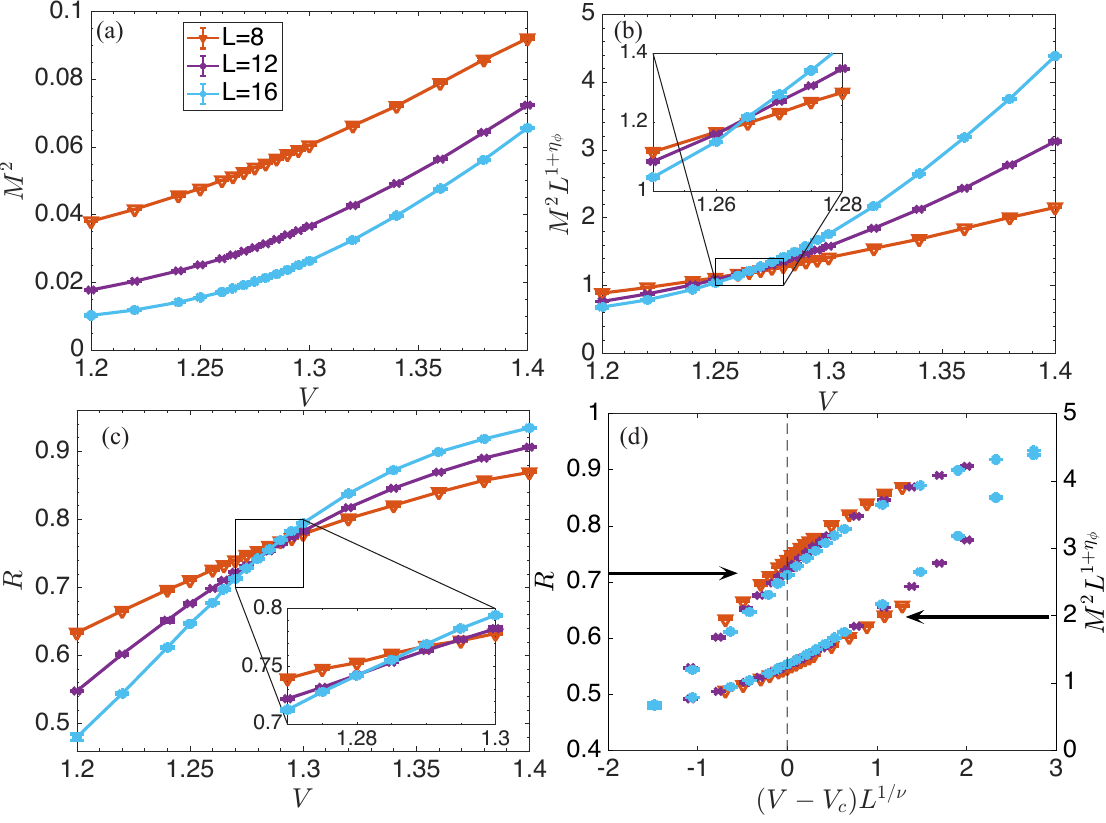}
\caption{\textbf{Observables of the $t-V$ model.} (a) Order parameter $M^2$, its (b) finite-size scaling crossing points and (c) correlation ratio $R$, and (d) data collapse with the two observables. Simulation is done with inverse temperature set to equal system size $\beta=L$. All four panels share the same legend in (d).}
\label{fig:fig_op}
\end{figure}

In finding the QCP of $t-V$ model, we perform the finite-temperature DQMC in various system sizes, setting $\beta$ to scale linearly with system size, since the fermion dispersion is linear ($z=1$) at the QCP. Fig.~\ref{fig:fig_op} shows the order parameter $M^2=S(0,0)$, where $S(k_x,k_y)=\sum_{i,j,\alpha}e^{-i(k_x,k_y)\cdot \vec{r_{ij}}}(n_{i,\alpha}n_{j,\alpha}-n_{i,\alpha}n_{j,\bar\alpha})$, correlation ratio $R=1-\frac{S(2\pi/L,0)}{S(0,0)}$, the finite-size crossing of them and data collapse of $M^2$ using the critical exponents from Ref.~\cite{Erramilli-Gross-2023}.  

Panel (b) shows crossing near $V=1.265$, drifting slightly towards larger $V$, which is a sign of small finite-size effect, while the crossing of correlation ratio, an RG invariant value by design, shows more severe drifting from $V=1.28$ towards smaller $V$. The data collapse of them, shown in panel (d), with the usual scaling form shows good collapse for different sizes with $V_c=1.27$.

\begin{figure}[htp!]
\includegraphics[width=\columnwidth]{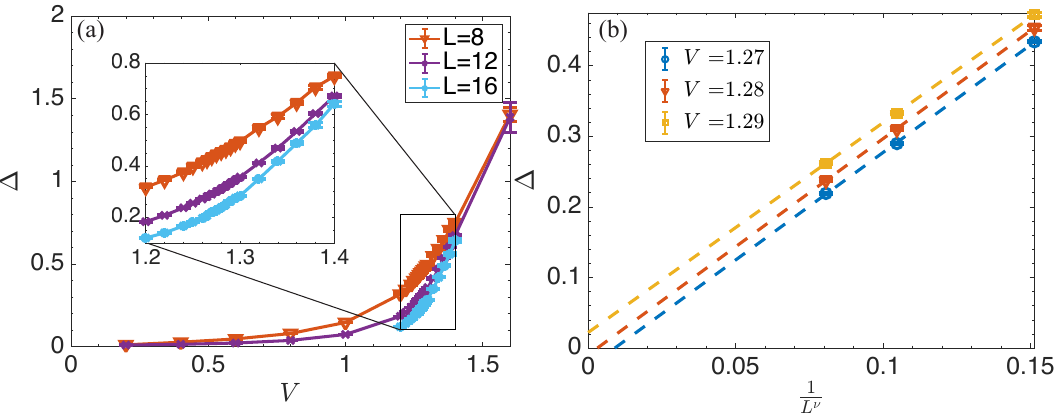}
\caption{\textbf{Single-particle gap versus $V$.} finite-size gap fitted using the decay of imaginary time Green's function $G(\vec{k},\tau)\propto e^{\Delta_\vec{k}}$ at Dirac points. (a) finite-size gap against $V$. (b) Scaling of finite-size gap against $1/L^\nu$. Dashed lines in panel (b) are straight lines fitted with the three largest system sizes.}
\label{fig:fig_gap}
\end{figure}

With the above setting, we also calculated the imaginary time decay of the Green's function $\tilde{G}(\vec{r}_i-\vec{r}_j,\tau)=\av{c_i(\tau)c^\dagger_j(0)}$, and its Fourier component $\tilde{G}(\vec{k},\tau)=\sum_{\vec{r}}e^{-i\vec{r}\cdot\vec{k}}\tilde{G}(\vec{r},\tau)$. At each $\vec{k}$ point, the Green's function decays exponentially as $\tilde{G}(\vec{k},\tau)=\av{c_k(\tau)c^\dagger_k(0)}\propto e^{\Delta_\vec{k}}$, which can therefore be used to calculate the single-particle gap at the Dirac points ($\vec{k}=(\pi,0)$ and $\vec{k}=(0,\pi)$).

As shown in Fig.~\ref{fig:fig_gap} (a), the gap remains 0 at small $V$ and opens at around $V=1.2$. Panel (b) shows the scaling of finite-size gap $\Delta$ against system size $L$ in the critical regime. At the QCP, the gap should close as $\Delta=1/L^{z\nu}$. The y-intercepts of the fitted straight dashed lines suggest that the gap opens near $V=1.27\sim 1.28$ in thermodynamic limit $L\rightarrow\infty$.

\begin{figure}[htp!]
\includegraphics[width=0.8\columnwidth]{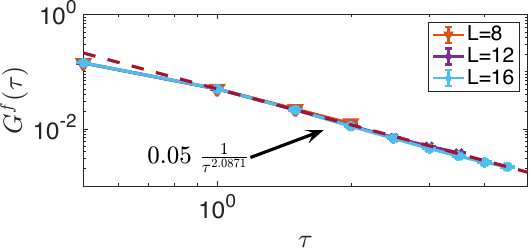}
\caption{\textbf{Imaginary time decay of the fermionic Green's function $\tilde{G}^f(\tau)$.} Dashed line is the power-law decay function with power $2+\eta_\psi$.}
\label{fig:fig_Gf_decay}
\end{figure}

For the onsite imaginary time Green's function $\tilde{G}^f(\tau)=\av{c_i(\tau)c^\dagger_i(0)}$, it should decay algebraically at the QCP with power $2+\eta_\psi$. As shown in Fig.~\ref{fig:fig_Gf_decay}, the decays for all system sizes coincide with the power-law decay with scaling dimension $\eta_\psi$ from Ref.~\cite{Erramilli-Gross-2023}.

\begin{figure}[htp!]
\includegraphics[width=\columnwidth]{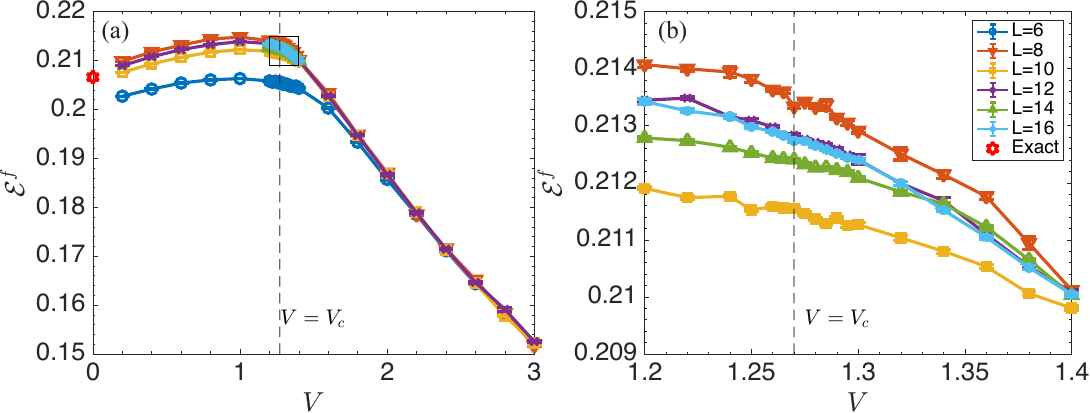}
\caption{\textbf{Nearest neighbor LN $\epsf$ against $V$.} In all cases, inverse temperature $\beta$ is set to equal to the system size. The red star in panel is the exact result of the free fermion model with system size and inverse temperature $L=\beta=1000$. Panel (b) is a zoom-in plot of the square region in (a). Panels (a) and (b) share the same legend in (a)}
\label{fig:fig_eps_V}
\end{figure}

Fig.~\ref{fig:fig_eps_V} shows the result of nearest neighbor $\epsf$ against repulsion strength $V$. For $V>V_c$, the system is in the gapped CDW phase, which explains the good convergence even in higher temperature. For $V<V_c$, the system is in the gapless DSM phase, and $\epsf$ is drifting up as $L$ and $\beta$ increase. The $L=\beta=12$ curve extrapolates to the ED result in $L=\beta=1000$, indicated by the red star in panel (a), thus showing good convergence to the ground state in thermodynamic limit. Panel (b) shows the zoom-in plot in the critical regime. The difference between different sizes only appears in the third decimal place. $\epsb$ peaks near the QCP slightly towards the DSM phase at around $V=1.2$.

\subsection*{$\epsf$ in finite temperature}

\begin{figure}[htp!]
\includegraphics[width=\columnwidth]{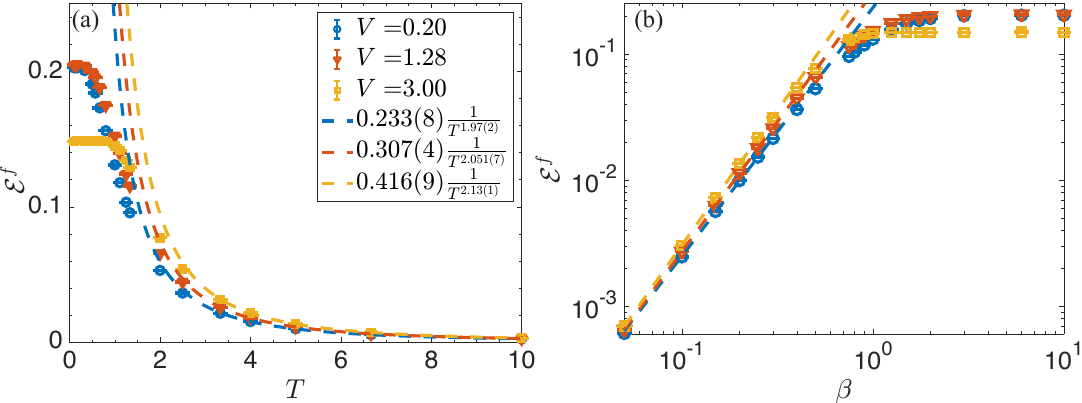}
\caption{\textbf{Nearest neighbor entanglement negativity $\epsf$ in finite temperature} The changes of $\epsf$ against (a) temperature $T$ and inverse temperature $\beta$. Dashed lines are fitted power-law decay curves at corresponding $V$. The two panels share the same legend in (a).}
\label{fig:fig_eps_T_fit}
\end{figure}

Fig.~\ref{fig:fig_eps_T_fit} shows the nearest neighbor LN $\epsf$ at finite temperature, with $V=0.20$ (DSM phase), $V=1.28$ (near QCP), and $V=3.00$ (CDW phase) respectively. The result of high-temperature fitting yields powers close to 2, which is the theoretical prediction in the free case~\cite{choi2023finite}.

\begin{figure}[htp!]
\includegraphics[width=\columnwidth]{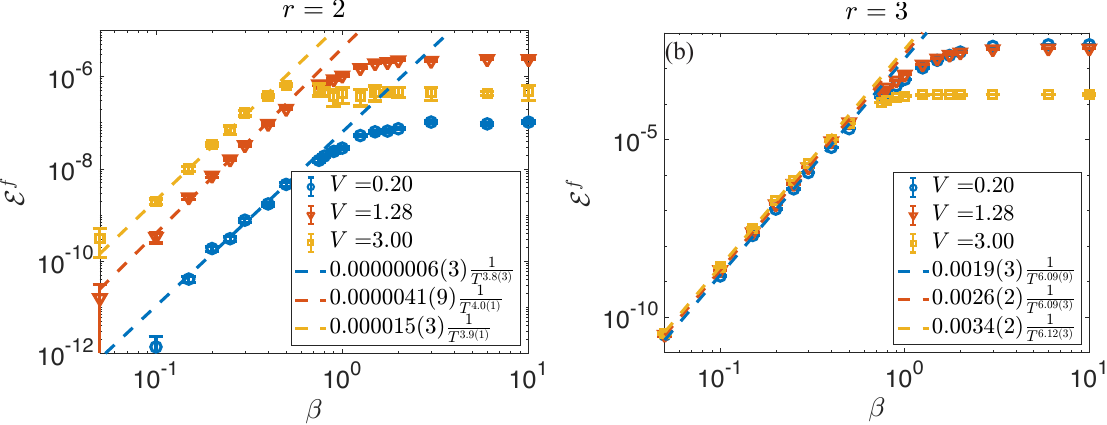}
\caption{\textbf{Separated entanglement negativity $\epsf$ in finite temperature} The changes of $\epsf$ against inverse temperature $\beta$, with $A_1$ and $A_2$ separated with bond-distance $r$ being (a) 3 and (b) 5. Dashed lines are fitted power-law decay curves at corresponding $V$.}
\label{fig:fig_eps_T_fit_r}
\end{figure}

If one further separate the two subregions, the power will increase as $2r$ as shown in the fitting in Fig.~\ref{fig:fig_eps_T_fit_r}, see Main text for the explanation. 


\begin{figure}[htp!]
\includegraphics[width=\columnwidth]{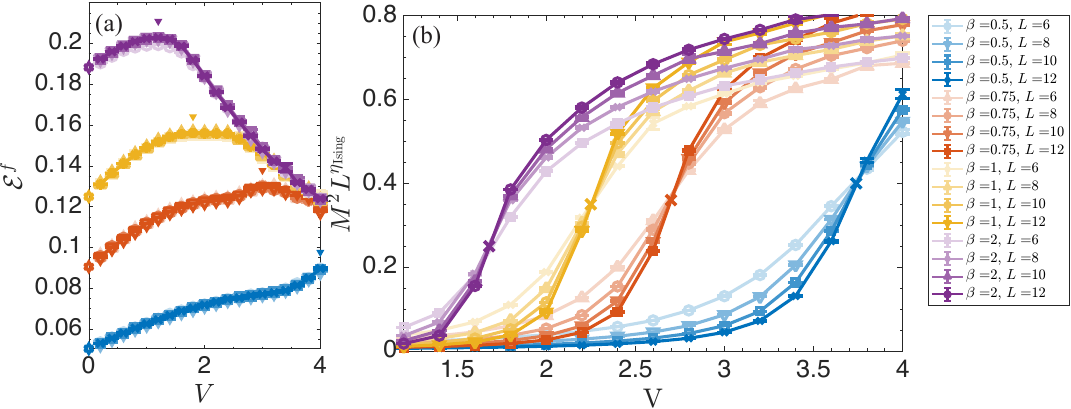}
\caption{\textbf{Entanglement negativity and finite-size scaling in different temperatures and system sizes.} (a) $\epsb$ with different system sizes in $\beta=0.75, 1, 2, 6$. The stars in different colors are the free fermion ED results done in $L=1000$ for corresponding temperatures. The triangles indicate the maxima of $\epsf$ along the $V$ axis (the purple dots in Fig.~\ref{fig:fig2} in the main text).(b) finite-size scaling of the order parameter $M^2$ against V. Crosses in different colors indicate crossing of the re-scaled observable in respective temperatures. The two panels share the same legend.}
\label{fig:fig_finite_T_size}
\end{figure}

To further ensure the $L=6$ finite-temperature data used to construct the heatmap in main text has converged to the thermodynamic limit, and to find the locations of finite-temperature phase transition $V_c$, we also simulate the system in larger system sizes in a few fixed temperature ($\beta=0.5, 0.75, 1, 2$).

As shown in Fig.~\ref{fig:fig_finite_T_size} (a), the finite-temperature $\epsf$ in these sizes has converged well, indicated by the overlapping of curves from different system sizes. For all temperatures, the curves extrapolate well to the free fermion (${V=0}$) ED result.

Fig.~\ref{fig:fig_finite_T_size} (b) shows the result of finite-size scaling with order parameter and the 2d Ising critical exponent $\eta_\text{Ising}=\frac{1}{4}$. The phase transition point, indicated by crosses in respective colors, makes up the finite-temperature phase boundary in the heatmap in main text.

\end{document}